# A Unified Approach to Attractor Reconstruction

Louis M. Pecora[1], Linda Moniz[2], Jonathan Nichols[3], and Thomas L. Carroll[1]

[1]*Code 6362, Naval Research Laboratory, Washington, DC 20375*

[2]*US Geological Survey, Patuxent Wildlife Research Center, Laurel, MD 20708, USA*

[3]*Code 5673, Naval Research Laboratory, Washington, DC 20375*

**What is the requirement for the determination of vector components for the reonstruction of an attractor from time series? This is a problem that has been studied for many years and everyone agrees that the problem consists of finding time delay, embedding dimension, and in the multivariate case which time series to use for each coordinate (although the latter is a much neglected problem). Most work has arbitrarily divided the problem into finding the delay and embedding dimension, separately. This division is the source of many problems. In addition, almost all approaches, and there are many, rely on heuristic methods or a choice of arbitrary physical scales (e.g. what constitutes a false neighbor) which are usually unknown. We view the construction of vectors for attractor reconstruction as a problem in finding a coordinate system to represent the dynamical state. What is mathematically necessary for any good coordinate system is that the coordinates be independent; this requirement is highlighted in Taken's theorem. To this end we develop a statistic to test for general, nonlinear functional dependence called the continuity statistic. This allows us to simultaneously test for delays and the necessity of adding more embedding dimensions with the same statistic. This helps us determine the best coordinates and gives us feedback on how well our**

**reconstruction is progressing. For chaotic systems we also need to know when we are taking delays that are too long. This latter situation is always mentioned, but never dealt with. We establish an undersampling statistic based on the geometric description of the attractor manifold when there are long delays. This latter statistic alerts us when we are using embedding parameters beyond what can be supported by the data. Together the continuity and undersampling statistic tell the practioner when he/she is doing the reconstruction right – something other statistics do not do.**

Abstract:

In the analysis of complex, nonlinear time series, scientists in a variety of disciplines have relied on a time delayed embedding of their data, i.e. attractor reconstruction. The process has focused primarily on heuristic and empirical arguments for selection of the key embedding parameters, delay and embedding dimension. This approach has left several long-standing, but common problems unresolved in which the standard approaches produce inferior results or give no guidance at all. We view the current reconstruction process as unnecessarily broken into separate problems. We propose an alternative approach that views the problem of choosing all embedding parameters as being one and the same problem addressable using a single statistical test formulated directly from the reconstruction theorems. This allows for varying time delays appropriate to the data and simultaneously helps decide on embedding dimension. A second new statistic, undersampling, acts as a check against overly long time delays and overly large embedding dimension. Our approach is more flexible than those currently used, but is more directly connected with the mathematical requirements of embedding. In addition, the statistics developed guide the user by allowing optimization and warning when embedding parameters are chosen beyond what the data can support. We demonstrate our approach on uni- and multivariate data, data possessing multiple time scales, and chaotic data. This unified approach resolves all the main issues in attractor reconstruction.

## I. Introduction

One of the most powerful analysis tools for investigating experimentally observed nonlinear systems is attractor reconstruction from time series which has been applied in many fields [1,2]. The problem of how to connect the phase space or state space vector **x**(*t*) of dynamical variables of the physical system to the time series *s*(*t*) measured in experiments was first addressed in 1980 by Packard et. al [3] who showed that it was possible to reconstruct a multidimensional state-space vector by using time delays (or advances which we write as positive delays) with the measured, scalar time series *s*(*t*). Thus, a surrogate vector $\mathbf{v}(t) = (s(t), s(t+t), s(t+2t),...)$ for **x**(*t*) could be formed from scalar measurements. This is essentially a particular choice of coordinates in which each component is a time-shifted version of the others with the time shift between adjacent coordinates the same.

Takens [4] and later Sauer et. al [5] put this idea on a mathematically sound footing by showing that given any time delay $\tau$ and a dimension $\Delta \geq 2$ box-counting dimension (**x**)+1 then, nearly all delay reconstructions are one to one and faithful (appropriately diffeomorphic) to the original state space vector **x**(*t*). These important theorems allow determination of system dynamical and geometric invariants from time series in principle.

The above theorems are existence proofs. They do not directly show how to get a suitable time delay $\tau$ or embedding dimension $\Delta$ from a finite time series. From the very start [3] emphasis has been put on heuristic reasoning rather than mathematically rigorous criteria for selecting statistics to determine $\tau$ and $\Delta$ and that remains true up to the present [2]. But many issues are unsettled. For example, there are no clear-cut statistical approaches for dealing with multiple time scales, multivariate data, and avoiding overly long time delays in chaotic data even though these problems are common in many systems and are often acknowledged as important issues in attractor reconstruction.

Various approaches to the $\tau$ problem have been autocorrelation, mutual information [6], attractor shape [7], and predictive statistics based on various models [8-10]. All these approaches have serious short comings (see Kantz and Schreiber [2] and Abarbanel [1] for critiques).

To determine $\Delta$ Kennel et al. [11,12] developed false nearest neighbor (FNN) statistics. These require one to choose an arbitrary threshold which ignores all structure under that scale which may be considerable for many attractors. Furthermore, in chaotic systems the statistic can be skewed by the divergence of nearby points on the attractor or the existence of two time scales so that the procedure may not truly terminate. That is, given a chaotic time series additional delays will eventually cause the FNN statistic to increase from trajectory divergences alone, leading to the incorrect addition of embedding dimensions. In addition, in time series with vastly different time scales the FNN statistic may achieve a minimum when the initial delays are equal to the fastest time scale incorrectly suggesting an embedding dimension lower than the true dimension.

Cao [13] has suggested a scale-free FNN approach, but this struggles with the fact that time-delay embeddings of any signal (even noise) will reach an asymptote at the same level as low-dimensional, deterministic systems. One should also compare Fraser's paper [14] which attempts to go beyond a two-dimensional statistic by using redundancies. The latter paper by its own admission has difficulties with computation with probability distributions when the embedding dimension increases.

The problem of a long time window in chaotic systems is often mentioned in passing, but only with simple admonitions that somehow the product $\tau \Delta$ should not get "too large." This is an acknowledgment of the above-mentioned FNN problem with diverging chaotic trajectories. Such problems point to the fact that until now there have not been statistics to check on overly long embedding times or to guide the user on the quality of the attractor reconstruction.

Finally, very little has been done with multivariate time series. Usually data analysis simply extends what is done for univariate time series [8,15] thus retaining the shortcomings. There are very few tests for optimal choices of time series to use from a multivariate set (outside of eliminating linear dependence using singular value decomposition [16]) or the Gamma test of Jones [17]). Time series with multiple time scales pose another difficult and unaddressed problem (an exception is the more recent work in Ref. [9], although those approaches rely on specific models).

We claim that the above approaches to attractor reconstruction artificially divide the problem into two problems thereby causing more difficulties than necessary while failing to address typical problems of multiple time series, overembedding [18,19] and multivariate data. In this paper we show that only one criterion is necessary for determining embedding parameters. This single criterion is the determination of functional relationships among the components of $\mathbf{v}(t)$ (the reconstruction vector). This allows us to find $\tau$ and $\Delta$ simultaneously and deal with the unsolved problems of multivariate data, excessively large $\tau \Delta$ and multiple time scales. For this reason we refer to our approach as a unified approach to attractor reconstruction.

## II. Continuity and Undersampling Statistics

Takens [4] (p.370, theorem 1(*iii*)) showed that, generically, $s(t+\tau)$ is functionally independent of $s(t)$. With this in mind we can provide a general requirement which quantifies the function independence of reconstruction coordinates. Consider a multivariate time series data set $\{s_i(t)|\ i=1,...,M\}$ sampled simultaneously at equally-spaced intervals, $t=1,...,N$. Then for each $\mathbf{v}(t)$ component we can inductively choose among $M$ time series and various delays (not necessarily equal). Thus, suppose we have a $\mathbf{v}(t)$ of $d$ dimensions $\mathbf{v}(t) = \left(s_{j_1}(t+\tau_1), s_{j_2}(t+\tau_2), ..., s_{j_d}(t+\tau_d)\right)$, where the $j_k \in \{1,...,M\}$ are various choices from the mutivariate data and, in general, each $\tau_k$ is different for each component; usually $\tau_1=0$. To decide if we need to add another component to $\mathbf{v}(t)$, i.e. increase the dimension $\Delta$ of the embedding space, we must test whether the new candidate component, say $s_{j_{d+1}}(t+\tau_{d+1})$ is *functionally independent* of the previous $d$ components. Mathematically we want,

$$s_{j_{d+1}}(t+\tau_{d+1}) \neq f\left(s_{j_1}(t+\tau_1), s_{j_2}(t+\tau_2), ..., s_{j_d}(t+\tau_d)\right) , \qquad (1)$$

for any function $f : \mathbf{R}^d \rightarrow \mathbf{R}^1$. Eq. (1) is the rigorous criterion for generating new $\mathbf{v}(t)$ components and is a general requirement for independent coordinates [20]. Using this approach we continue to add components to $\mathbf{v}(t)$ until all possible candidates for new components from all time series and for all $\tau$ values are functions of the previous components. In this way we have found $\Delta$ and all delays *simultaneously* and are done. Note, we have not separated the $\tau$ and $\Delta$ problems – they are found together. Below we develop a statistic to select $s_{j_{d+1}}$ and $\tau_{d+1}$ that fulfill the independence criterion.

We build our test statistic for functional relationships (Eq. (1)) on the simple property of continuity. It captures the idea of a function mapping nearby points in the domain to nearby points in the range yet assumes nothing more about the function. We use a version of a continuity statistic [21,22] with a new null hypothesis [23] that is more flexible in the presence of noise.

For example, suppose $\mathbf{v}(t)$ has $d$ coordinates and we want to test if we need to add another $d+1$st component. Eq. (1) is a test for evidence of a continuous mapping $\mathbf{R}^d \rightarrow \mathbf{R}^1$ (Eq. (1)). Following the definition of continuity we choose a positive $\varepsilon$ around a fiducial point $s_{j_{d+1}}(t_0 + \tau)$ in $\mathbf{R}^1$ ($\tau$ is fixed for now). We pick a $\delta$ around the fiducial point $\mathbf{v}(t_0) = \left(s_{j_1}(t_0 + \tau_1), ..., s_{j_d}(t_0 + \tau_d)\right)$ in $\mathbf{R}^d$ corresponding to $s_{j_{d+1}}(t_0 + \tau)$. Suppose there are $k$ points in the domain $\delta$ set. Of these suppose $l$ points land in the range $\varepsilon$ set. We invoke the continuity null hypothesis that those $l$ points landed in the $\varepsilon$ set by chance with probability $p$. This is shown schematically in Figure 1. A good choice for data which we use here is $p=0.5$, i.e. a coin flip on whether the points are mapped from $\delta$ into $\varepsilon$. Other choices are possible [21,23], but $p=0.5$ is actually a standard null, harder to reject and very robust under additive noise. Now we pick the confidence level $\alpha$ at which we reject the null hypothesis. If the probability of getting $l$ or more points in the $\varepsilon$ set (a binomial distribution) is less than $\alpha = 0.05$, we reject the null. Other confidence levels are possible and should be considered if one can do a risk analysis, but here we stay with the 0.05 statistics standard to insure that there are at least 5 points in both $\delta$ and $\varepsilon$ balls (we comment more on this in the conclusions). We note here that an essential part of the reconstruction process is to report not only delays and embedding dimension, but the null hypotheses and the confidence levels used. We do this throughout.

Repeat this process by decreasing $\varepsilon$ and varying $\delta$ until we cannot reject the null. Call the smallest scale at which we can reject the null our **continuity statistic** $\varepsilon^*$. We sample the data sets at many fiducial points and calculate the average $<\varepsilon^*>$. The data determines $\varepsilon^*$, the smallest scale for which we can claim a function relationship. If the time series is too short to support an embedding, this will be reflected in large and uniform $\varepsilon^*$ for any delay.

We suggest that a way to view this statistic is analogous to how we look upon tests for linear independence. If we examine normalized covariances of separate time series, we rarely see either 1's or 0's, but rather numbers between 0 and 1. The closer to 0 a covariance is the less linearly dependent are the two associated time series. In this more general case, the larger <$\varepsilon$*> the more functionally independent are two coordinates.

It is the reliance on statistics calculated from the data to provide a length scale <$\varepsilon$*> rather than imposing one arbitrarily (e.g. as in FNN approaches) which enables the continuity statistic to guide us in determining the reconstruction parameters and in judging the quality of our reconstruction. If we succeed in reducing <$\varepsilon$*> by adding components at proper delays we will know we are doing well in reconstructing the attractor. When we can no longer reduce <$\varepsilon$*>, we are done – this is the best we can do with the given data set.

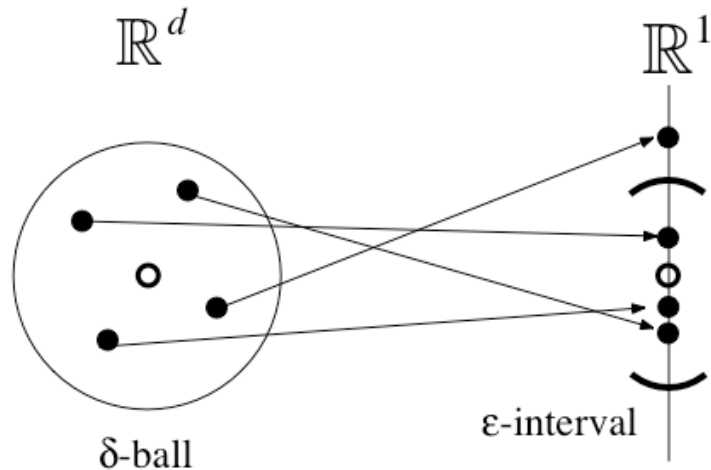

Figure 1. The $\delta$ and $\varepsilon$ sets and their data points in $R^d$ and $R^1$ respectively. Of the $k$ points ($k$=4, here) in the $\delta$-ball $l$ ($l$=3, here) are mapped into the $\varepsilon$ interval and 1 is mapped outside. The probability of interest by which we accept or reject the null hypothesis is the cumulative binomial for getting 3 or more points in the $\varepsilon$ interval with probability $p$ (see text).

Applying the continuity test inductively, we start with one component and build up $\mathbf{v}(t)$ one dimension at a time by examining the continuity statistic $<\varepsilon^*>$ as a function of delay and/or data set. If possible we choose $\tau$ at a local maximum of $<\varepsilon^*>$ to assure the most independent coordinates (as in Eq. (1)). If $<\varepsilon^*>$ remains small out to large $\tau$, we need not add more components; we are done and $\Delta = d$. In the multivariate case we can extend this to generating an $<\varepsilon^*>$ for each potential new component $s_j(t+\tau), j = 1,...,M$ of $\mathbf{v}(t)$ and choose that component which has the least functional relationship (a maximum of $<\varepsilon^*>$) to the previous $d$ components.

In principle the continuity statistic should be the only test we need since it not only determines a good set of $\mathbf{v}(t)$ components, but also a "stopping point" when there are no more independent components, hence automatically giving $\Delta$ and $\tau$ together. But real data is finite in number and resolution. Because one cannot get nearest neighbors arbitrarily close for finite data, eventually, with large enough delays any two reconstruction vectors from a chaotic time series will necessarily have some components appear randomly scattered on the attractor. This phenomenon is endemic to all chaotic systems. Mathematically the problem is that in looking at points distant in time we are essentially looking at points distributed on a high iteration of the flow or map. This indicates the manifold of the system is very folded and contorted and we are undersampling it. Thus, one cannot discern a loss in functional relation from an undersampled manifold both of which give a large $<\varepsilon^*>$. It is easy to see this effect in a simple one-dimensional map like the logistic map. Kantz and Olbrich [18] showed a similar overembedding phenomenon for simple maps in high dimensions.

We developed an **undersampling statistic** to detect when $\mathbf{v}(t)$ components enter this undersampling regime. We use the null hypothesis that at least one of the $\mathbf{v}(t)$ components is randomly distributed on the attractor. We test to see if component-wise difference between a point $\mathbf{v}(t_0)$ and its nearest neighbor $\mathbf{v}_{NN}$ in $\mathbf{R}^{d+1}$ (the combined $\delta$ and $\varepsilon$ spaces) is on the order of typical distances between randomly distributed points on the attractor. To do this test we generate the baseline probability distribution of differences between randomly chosen points from each time series. It is easy to derive the distribution $\sigma_j(\xi)$ of differences $\xi$ between randomly chosen points from the probability distribution $\rho_j(x)$ of values in the time series. We calcuate $\rho_j(x)$ for each time series $j = 1,...,M$ (e.g. by binning the data). Then it is easy to show under the assumptions of randomly chosen points that $\sigma(\xi) = \int \rho(x)\rho(\xi - x)\,dx$. Then for each component $i$ of the difference between a point $\mathbf{v}(t_0)$ and its nearest neighbor $\mathbf{v}_{NN} - \mathbf{v}(t_0)$ the probability of getting a value less than or equal to $\xi_i = \left|(\mathbf{v}_{NN} - \mathbf{v}(t_0))_i\right|$ (the NN distance for the $i$th component) at random from the $j_i$th time series is $\Gamma_i = \int_0^{\xi_i} \sigma_{j_i}(\zeta)d\zeta$. Let $\Gamma = \max\{\Gamma_i\}$. As with the continuity statistic we average $\Gamma$ over fiducial points on the reconstruction.

    Choosing the confidence level $\beta$ for rejection or acceptance of the undersampling null hypothesis presents a separate issue and we would like to comment on it here. Again, we could choose the standard level for rejection at $\beta=0.05$, so that if $\Gamma \le 0.05$ we reject the null hypothesis and accept that the time delays are not too large so that the average distances do *not* appear to be distributed in a random fashion. We do this at times in the following examples, but we also display another approach that is an advantage of our unified method. This is to monitor $\Gamma$ vs. $\tau$ as we add reconstruction components (dimensions) and when $\Gamma$ drops precipitiously we stop. The main point is that we report not only on the reconstruction parameters ($\Delta, \tau_i$, and, for multivariate time series which components were used for the reconstruction vector), but also what the null hypothesis rejection levels are. Thus, we might say we have a two-dimensional reconstruction that uses time series #1 at a delay of 0 and time series #2 at a delay of 15 with an $\alpha=0.05$ and a $\beta=0.1$. Stating the level of confidence allows quantative judgement by others of the

quality of the reconstruction which, in a certain sense, is also a statement about the quality and quantity of the data. In many cases there are no absolute threshold values so it is fitting that the problem be cast in a hypothesis-testing framework. We remark more on this in the conclusions.

Note that for all our statistics our data points are gathered using a temporal exclusion window [24] or using strands in place of points [11] to avoid temporal correlations.

## III. Applications

We now present the continuity statistic $<\varepsilon^*>$ and the undersampling statistic $\Gamma$ for some typical systems of increasing complexity using univariate and multivariate data. Our notation for labeling different $<\varepsilon^*>$ results for Eq. (1) is to simply label the current delays in $\mathbf{v}(t)$. So 0,10 labels the $<\varepsilon^*(\tau)>$ which is testing for a functional relation from the two-dimensional $\mathbf{v}(t)$ (the time series and a 10-step advanced shifted version) to the next possible third-dimensional component time shifted by $\tau$. All time series are normalized to zero mean and standard deviation of 1.

L. Pecora and , Naval Research Laboratory

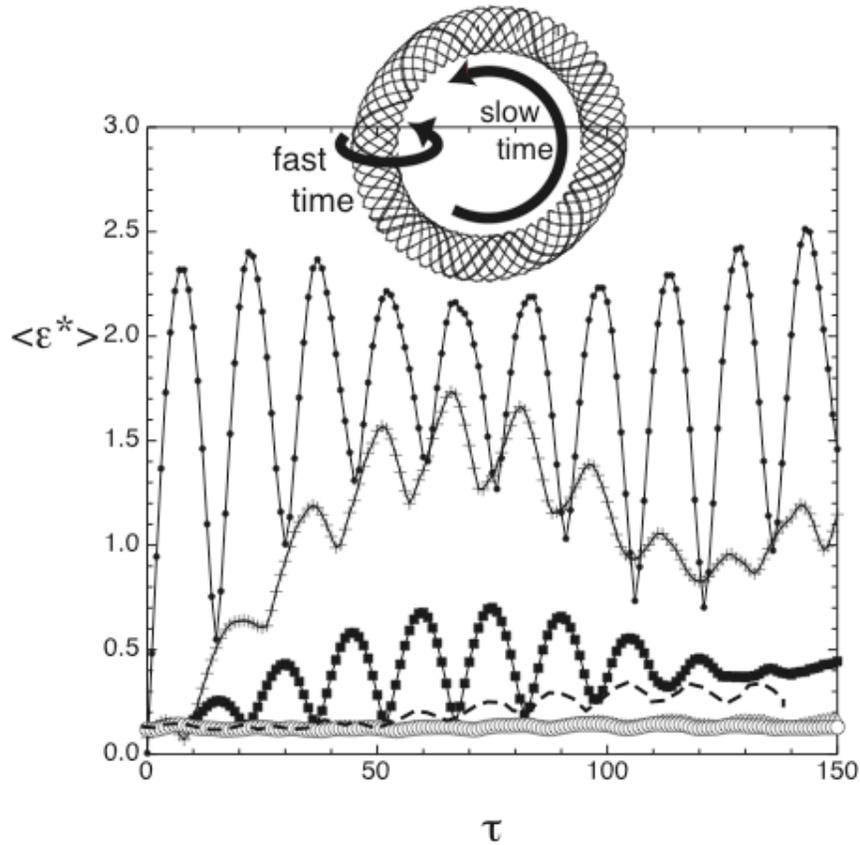

Figure 2. Continuity statistic for a quasiperiodic system. Inset: The torus for the quasiperiodic system with two time scales. The continuity statistic for the quasiperiodic system for the embeddings 0 (—●—); 0, 8 (—+—); 0, 8, 67 (—■—); 0, 8, 67, 75 (—▽—); and 0, 8, 67, 75, 112 (—○—); and the constant $\tau = 8$ embedding (_ _ _) using 8 dimensions.

### A. Quasiperiodic, Multiple Time-Scale System

A non-trivial case is a quasiperiodic system with different time scales. Fig. 2 inset shows the torus of this system. The slow and fast times are in the ratio of $2.5\pi:1$ (approximately 8:1) and the time series is sampled at 32 points per fast cycle. The continuity statistic shows only four dimensions are needed since $<\varepsilon^*>$ falls to and remains at a low level after adding the $\tau_4=75$ component. This is correct since any 2-Torus can be embedded in four dimensions or less. The two time scales are correctly captured in the ratio of $\tau_2:\tau_3$ . In comparison, using the standard constant delay from the first minimum of the mutual information takes more than 8 dimensions to embed the torus.

### B. Lorenz Attractor Reconstruction

We tested the unified approach on a chaotic three-dimensional Lorenz system [25] with $\sigma=10$, $b=8/3$, $\rho=60$. The *x*-time series was generated using a 4th-order Runge-Kutta solver with a time step of 0.02 and 64,000 points. We calculated $<\varepsilon^*>$ by averaging $\varepsilon^*$ over 500 random points on the reconstructions. The results are shown in Fig. 3. This system is chaotic with $<\varepsilon^*>$ eventually increasing with $\tau$ because of undersampling so we add the undersampling statistic, $\Gamma$, at the bottom of the figure.

A three-dimensional $\mathbf{v}(t)$ significantly lowers the $<\varepsilon^*>$ value out to near 300 time steps where it begins to rise. If we try to add another $\mathbf{v}(t)$ component, for example, with a $\tau=350$, the undersampling statistic rises drastically and goes above the 5% confidence level indicating that the increases in $<\varepsilon^*>$ result from a folded manifold that is undersampled. Nonetheless, the time window over which we have a good embedding is rather wide, about 300 time steps. Using the first minimum in the mutual information ($\tau\approx16$) would require a $\mathbf{v}(t)$ of about 16 components to accomplish the same reconstruction with constant $\tau$ embeddings. The continuity statistic provides confidence that details in the attractor down to a scale of 0.2 attractor standard deviation is real. This detail can be lost using arbitrary thresholds of other tests (e.g. mutual information and FNN). The dashed line shows the effect of gaussian white noise of 16% of the time-series standard deviation added to the data. Despite this high noise level much of the $<\varepsilon^*>$ structure remains. This noise robustness comes from our choice of the $p=0.5$ probability for the $<\varepsilon^*>$ null hypothesis. Similar results occur from shorter time series.

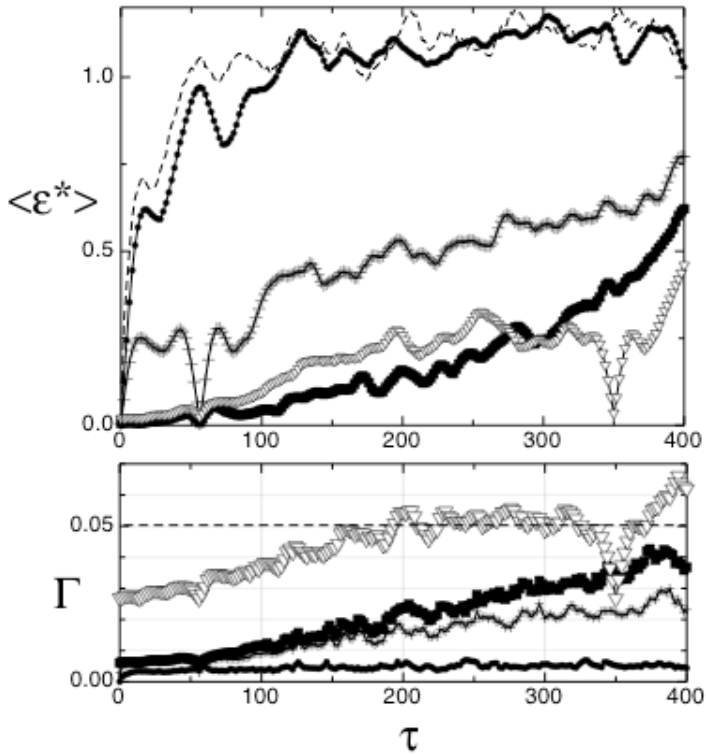

Figure 3. Continuity $<\varepsilon^*>$ and undersampling $\Gamma$ statistics for Lorenz time series plus $<\varepsilon^*>$ when 16% gaussian white noise is added to time series (- - - -). Advances: 0 (—●—); 0, 56 (—+—); 0, 56, 14 (—■—); 0, 56, 14, 350 (—▽—).

### C. A Multivariate Test Case

For the case of multivariate time series using all three of the Lorenz system's components, $x, y,$ and $z$ we calculated $<\varepsilon^*>$ for all possible combinations of components allowing for time delays or advances (±100 time steps). The results imply a surprising conclusion. The best set of components for the three-dimensional **v**(*t*) is (*x*(0), *x*(17), *z*(0)), i.e. no *y* component. The <$\varepsilon^*$> values for the *y* time series are below those of the *x* time series at almost all delay values. On reflection we realize that there is no reason why the variables of the physical equations of motion are best for attractor reconstruction.

### D. An Optimization Approach to the Reconstruction

Thus far we have taken the simple direct route of determining the delays sequentially by choosing relative maxima. This can be characterized as a form of greedy algorithm [26]. A more general and effective way to calculate the proper delays for a time series is to embed the signal in *d* dimensions, and choose all *d* delays simultaneously by numerical optimization. In this optimization procedure, the time series is embedded in *d* dimensions, and the statistic <$\varepsilon^*$> is calculated for the next *d+1*st component at each delay from 1 to 100. Since the goal of the unified approach is to minimize the continuity statistic we average the <$\varepsilon^*$> values for the 100 delays. This delay-averaged quantity <<$\varepsilon^*$>> can be thought of as the average continuity of the *d* dimensional embedding to the next (*d+1*st) component. This calcution is then done using various *d* values while monitoring the undersampling statistic.

We chose a downhill simplex method, with simulated annealing [27] (there were many local minima) for the numerical minimization. In order to speed the calculation, rather than vary the radius $\delta$ of the *d* dimensional embedding in $\mathbf{R}^d$, a fixed number of neighboring points on the *d* dimensional embedding was chosen (40 in this case). From the binomial distribution, if there is a 50%, probability of any particular point landing in the $\varepsilon$ set, there is a 5% chance of having 25 points all land in the $\varepsilon$ set, so in this routine, $\varepsilon$ is the radius on the $d+1^{st}$ component which contains 25 of the 40 points from the $\delta$ neighborhood on the *d* dimensional embedding.

L. Pecora and , Naval Research Laboratory

The minimum value of $<<\varepsilon*>>$ was calculated from a 30,000 point time series from the Lorenz system, with a time step of 0.03. For the calculation, 10,000 fiducial points were randomly chosen from the 30,000 point time series. For a 3-dimensional embedding, the minimum $<\varepsilon*>$ was 0.561, at delays of (0, 54, 34). For a 4-dimensional embedding, the minimum $<\varepsilon*>$ was 0.510, at delays of (0, 52, 35, 68). The optimized approach does show that adding a fourth component is probably warranted although the improvement in the reconstruction is not big. We can compare these with our greedy algorithm given above by multiplying the latter's delays by 2/3 since the Runge-Kutta time step for the latter was 0.02 while here we used 0.03. This gives (integer) delays of (0, 37, 9) in step sizes of 0.03. One pair of delays (37 and 35) agree, but the greedy algorithm resulted in the addition of a short delay whereas the optimized version suggests adding components to the reconstruction vector using larger delays. There were actually many local minima seen using the optimization method, so it is not surprising that different algorithms give different results. There is more than one "good" set of delays for reconstruction, just as there is more than one "bad" set of delays. The optimization approach is more rigorous, but simply choosing local maxima on the $<\varepsilon*>$ plot works almost as well if there are obvious maxima.

**E. Neuronal Data**

We applied these statistics to neuronal data taken from a lobster's stomatogastric ganglia [28] in Figure 4. These neurons are part of a central pattern generator and exhibit synchronized bursting similar to Hindmarsh and Rose models [28,29]. We find that the time delays ($\tau$= 0, 278, 110, 214, 56) range from 56 to 278, a factor of five. Five dimensions are required for a reconstruction at an undersampling statistic confidence level of $\beta$=0.1. This agrees qualitatively with a detailed model of the ganglia by Falcke et. al [30]. Using a constant delay established by the 1st minimum of the mutual information

17.    10:24 AM, June 30, 2006

($\tau$=32) five

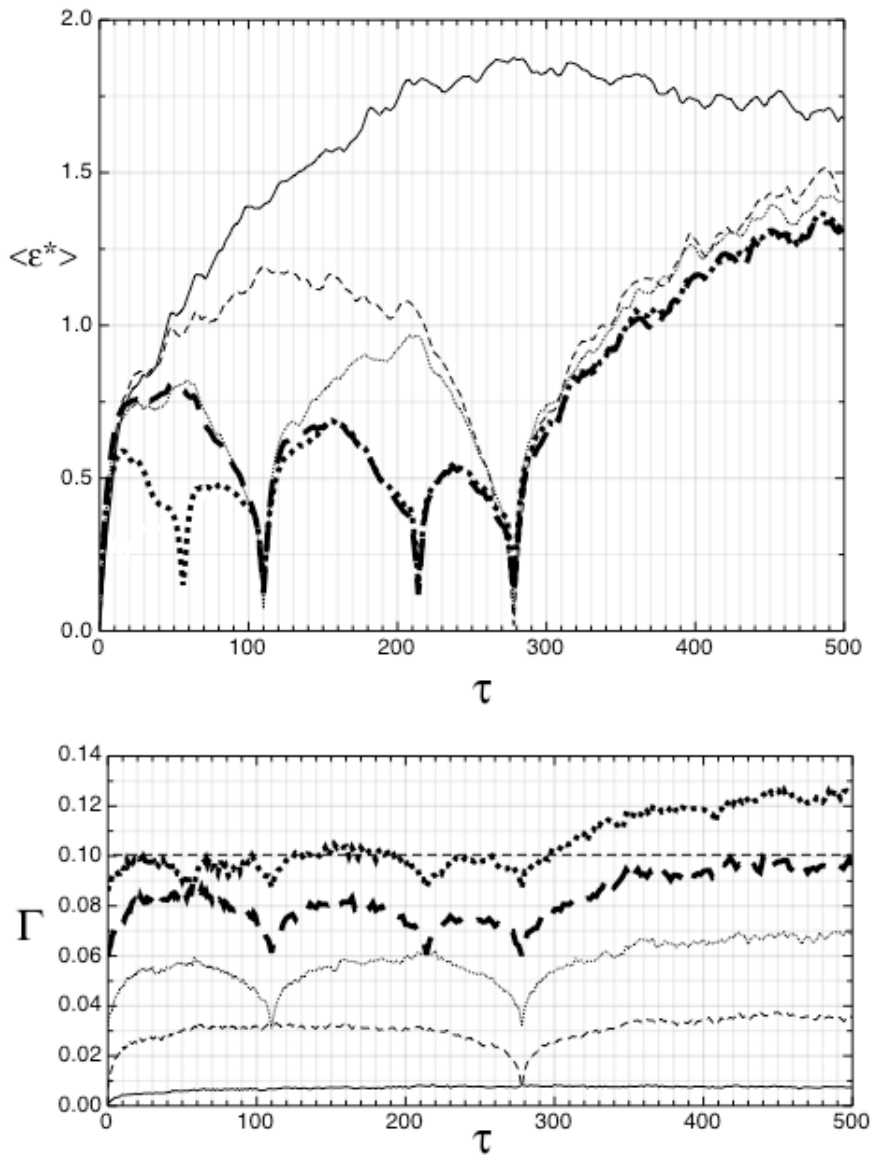

Figure 4. Continuity $<\varepsilon^*>$ and undersampling $\Gamma$ statistics for Lobster time series. Advances: 0 (———); 0, 278 (— — — —); 0, 278, 110 (⋯⋯⋯⋯); 0, 278, 110, 214 (■ ■ ■ ■); 0, 278, 110, 214, 56 (▬▬▬▬▬).

dimensions still does not capture the long time scales beyond $\tau$=160, whereas we have structure out to near $\tau$=300. Adding larger delays causes the undersampling statistic to go much higher than 0.1 and we decided to stop with the five delays we have, but we always note what confidence level we used.

## IV. Conclusions

Our unified approach uses statistics which faithfully adhere to the rigorous mathematical conditions of the Takens' embedding theorem ($<\varepsilon^*>$) and the accurate geometric picture of excessive time delays ($\Gamma$). Both statistics give good indications when we are constructing a good reconstruction ($<\varepsilon^*>$) and when we have gone beyond what the data will allow ($\Gamma$), e.g. overembedding. Hence, unlike other reconstruction approaches we have an indication of the quality of the embedding at a given level of confidence.

As we mentioned the $<\varepsilon^*>$ statistic is a generalization of the concept of linear correlation, but note that it is asymmetric in general. For example, given two time series, say $x(t)$ and $y(t)$, we might get very small $<\varepsilon^*>$ for the test of functionality $x(t) \rightarrow y(t)$ indicating $y(t)$ is a function of $x(t)$, but get a large $<\varepsilon^*>$ for the reverse relationship $y(t) \rightarrow x(t)$ indicating that $x(t)$ is not a function of $y(t)$. A simple example is $y(t)=x^2(t)$.

The statistics $<\varepsilon^*>$ and $\Gamma$ are data dependent as all statistics should be. Adding more time series points will lower the continuity statistic and improve the undersampling statistic so that the acceptable time window for an embedding will enlarge and more $\mathbf{v}(t)$ components will not be needed. The statistics replace arbitrary time and length thresholds with probabilities so that physical scales (which are often unknown) are derived and data-dependent, but not assumed at the outset.

The data-dependent method has the advantage (over current methods) of yielding unambiguous information about the ability to embed a data set at all. If it is not possible to choose successive delays so that the continuity statistics fall, the data set does not include enough information to produce independent embedding coordinates.

We have chosen "standard" statistical confidence levels here of $\alpha=0.05$ and $\beta=0.05$ (2 sigmas) in some cases. In others we stop the reconstruction when the undersampling statistic falls precipitously and give the reconstruction parameters along with the level for null hypothesis rejection.  In all problems with statistics choosing the level for the null hypothesis rejection is a separate problem and usually depends on each particular situation.  Finding the best confidence level would involve doing something like a risk analysis for which payoffs and penalties for correctly or incorrectly rejecting the null hypothesis are known.  Whether such an analysis is or is not possible (and often it is not), we believe that these levels should always be give so others can judge the quality of the reconstruction.  Because our statistics do not involve arbitrary scales, but rather null hypothesis we can always give the probability that the reconstruction is right.

Finally, we note that the problem of observability for delay embeddings might be addressed with our statistic since in this case there is no closed form solution for chaotic systems [31].

We would like to acknowledge helpful conversations with M. Kennel, L. Tsimring, H.D.I. Abarbanel, and M.A. Harrison.